\documentclass[aps,prd,twocolumn, preprintnumbers, floatfix, superscriptaddress,showpacs,nofootinbib,showkeys]{revtex4}
\usepackage[dvips]{graphics}
\usepackage{amsmath,amsfonts,bm}
\usepackage{epsfig,bm}
\usepackage{graphicx}
\def\MeijG[#1][#2][#3][#4][#5][#6]{G^{#1}_{#2}\left(#3,#4\left|\begin{array}{c}#5\\#6\end{array}\right|\right)}

\begin{document}

\preprint{LMU-ASC 16/13}

\affiliation{Ludwig-Maximilians-Universit\"at M\"unchen, Fakult\"at f\"ur Physik, Arnold Sommerfeld Center for Theoretical Physics, D–80333 M\"unchen, Germany}       

\author{Oscar Cat\`a}
\affiliation{Ludwig-Maximilians-Universit\"at M\"unchen, Fakult\"at f\"ur Physik, Arnold Sommerfeld Center for Theoretical Physics, D–80333 M\"unchen, Germany}       

\title{Revisiting $ZZ$ and $\gamma Z$ production with effective field theories}
\begin{abstract}
A complete and systematic effective field theory analysis of new physics effects in ${\bar{f}}f\to ZZ$ and ${\bar{f}}f\to \gamma Z$ is performed. Results are presented for the different initial and final-state polarized differential cross sections in terms of oblique, gauge-fermion and neutral triple gauge corrections (nTGC). Phenomenological signatures for new physics detection at the LHC and at future linear colliders are discussed. In comparison to $WW$ production, they follow a completely different pattern: nTGC only appear at NNLO in the effective field theory expansion and, accordingly, are extremely suppressed. However, in the high energy regime, $v\ll\sqrt{s}\ll 3$ TeV, nTGC are shown to neatly dominate the longitudinal-transversal final-state polarizations over the standard model background. Their tiny effects get hugely amplified at energies $\sqrt{s}\sim (0.6-1)$ TeV and can easily generate up to 20\% deviations over the standard model predictions. 
\end{abstract}

\maketitle

\section{Introduction}\label{secI}

Gauge boson pair production has become one of the main tools to test the standard model structure of the gauge boson self-couplings and place bounds on new physics corrections~\cite{Gaemers:1978hg,Hagiwara:1986vm}. Starting at LEP~\cite{Schael:2013ita}, the scrutiny of triple gauge vertices has become a standard strategy to probe new physics in the pure gauge sector and holds a prominent position in the physics programs at the Tevatron, the LHC and future linear collider facilities. Such analyses rely on the following points: (i) the triple gauge vertex corrections are assumed to be the dominant sources of new physics in ${\bar{f}}f\to WW,\gamma Z, ZZ$, for instance over gauge-fermion corrections, which are neglected; and (ii) the kinematic invariants that parametrize the triple gauge vertices are upgraded to form factors in order to preserve unitarity at high energies. (See~\cite{Gounaris:1996rz} for a general review. For concrete studies at hadron colliders, the reader is referred to~\cite{Baur:1992cd,Ellison:1998uy,Gounaris:1999kf,Baur:2000ae}.)    

There are some aspects of the standard analysis above that are not entirely satisfactory. First of all, gauge-fermion operators are parametrically on the same footing as the triple gauge operators for charged pair production and even dominant for neutral diboson production. Moreover, they can be related to triple gauge and oblique operators through field redefinitions, and neglecting them is in general inconsistent~\cite{Buchalla:2013wpa}. Gauge-fermion effects could still be numerically suppressed, but so far the LEP global electroweak fit gives no indications thereof: constraints on gauge-fermion and triple gauge operators are comparable~\cite{Han:2004az}. 

A form factor ansatz for the triple gauge vertex kinematical invariants is also objectable. Triple gauge parameters encode the underlying dynamical content of the electroweak theory. Regardless of the specific form of this UV-complete theory, at low energies it has to satisfy the $SU(2)\times U(1)$ standard model symmetry. The form factor ansatz is however blind to the standard model symmetry.

Both points can be improved if one adopts an effective field theory approach. To build an effective field theory one only needs to specify the field content of the theory and the symmetries to be obeyed at a given scale and endow the resulting set of operators with a power-counting, such that they are arranged as a series in inverse powers of a new physics scale $\Lambda$. Such an approach is by construction model-independent, $SU(2)\times U(1)$ invariant, and systematically improvable. Moreover, since the theory is only valid up to the scale $\Lambda$, it is automatically protected against unitarity breaking. The power-counting provides an ordering criteria for the operators and thus selects the most relevant features at a given energy scale, filtering out the unnecessary details. As a result, triple gauge vertex corrections can be expressed in terms of a small set of coefficients while respecting the standard model gauge symmetry.    

Effective field theory methods were applied in the past to the study of both charged and neutral gauge boson pair production. However, a complete and systematic analysis for $WW$ production was only recently performed in~\cite{Buchalla:2013wpa}. There it was shown that, once a complete set of effective operators is included, the leading new physics corrections to $WW$ production at LHC and linear colliders can be parametrized entirely in terms of gauge-fermion operators.    

In this paper I will extend the previous analysis to neutral gauge boson $ZZ$ and $\gamma Z$ production by working out the full set of new physics contributions affecting ${\bar{f}}f \to ZZ$ and ${\bar{f}}f\to \gamma Z$ and studying their impact at energy scales $s\gg v^2$. Fast convergence of the EFT expansion additionally requires that $s\ll \Lambda^2$. Given that new physics effects (if present) are expected to appear at the TeV scale, in practice this entails a rather broad energy window, namely $\sqrt{s}\sim (0.6-1)$ TeV, which nicely fits the operational energies of the LHC and future linear colliders. 

Since the nature of the scalar sector of the standard model is still unclear, it is advisable to work both with a linear and  a nonlinear realization of electroweak symmetry breaking. The linear realization assumes that the dynamics of electroweak symmetry breaking is weakly coupled, while the nonlinear realization is especially suited for strongly-coupled scenarios. In the former case, new physics is decoupled and $\Lambda$ is a free parameter; in the latter instead one naturally expects $\Lambda\sim 4\pi v\simeq 3$ TeV. 

Detection of new physics effects in neutral diboson production is a rather challenging task. Unlike the charged case, neutral triple gauge vertex effects are extremely subtle: the standard model contribution cancels at tree level and only appears at one-loop as anomalous fermion triangles, which bring ${\cal{O}}(10^{-4})$ contributions~\cite{Renard:1981es,Barroso:1984re,Gounaris:2000tb} to some of the triple gauge parameters. New physics effects only appear as NNLO effective operators and are likewise suppressed. As a result, the cross section for both $\gamma Z$ and $ZZ$ production is overwhelmingly dominated by tree level contributions in the $t$ and $u$ channels. 

In this work I will show that a final-state polarization analysis is the right tool to unveil new physics in $ZZ$ and $\gamma Z$ production.  This is in sharp contrast to the charged case. The main observation is that standard model physics  predominantly affect the final-state transversal (TT) polarizations. Anomalous nTGV effects are instead dominant for LT polarizations, where the standard model is only present through one-loop effects, which quickly decouple as $\ln s/s^2$~\cite{Gounaris:2000tb} to comply with anomaly consistency conditions. Therefore, even though LT polarizations are largely suppressed, they offer a remarkably clean test of anomalous nTGC: despite the smallness of the coefficients, at $\sqrt{s}=(0.6-1)$ TeV new physics corrections get extremely magnified, typically to ${\cal{O}}(20\%)$ corrections. 

This paper will be organized as follows: in Section~\ref{sec:II} I will discuss the most general form of new physics corrections in diboson production both at linear and hadron colliders. In Section~\ref{sec:III} I will work out the relevant effective field theory operators both in the linear and nonlinear realizations and give expressions for the new physics corrections in terms of the EFT coefficients. Section~\ref{sec:IV} collects the results for the (initial and final-state) polarized cross sections for both $e^+e^-\to ZZ$ and $e^+e^-\to \gamma Z$ processes. In Section~\ref{sec:V}, I compare the results obtained with those of $WW$ production. Conclusions are given in Section~\ref{sec:VI}.


\section{New physics effects in $ZZ$ and $\gamma Z$ production}\label{sec:II}
 
\begin{figure*}[t]
\begin{center}
\includegraphics[width=3.2cm]{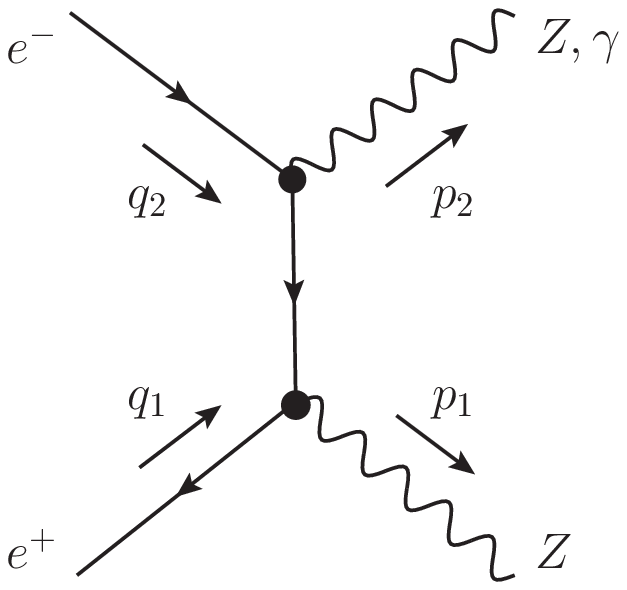}
\hspace{0.5cm}
\includegraphics[width=3.2cm]{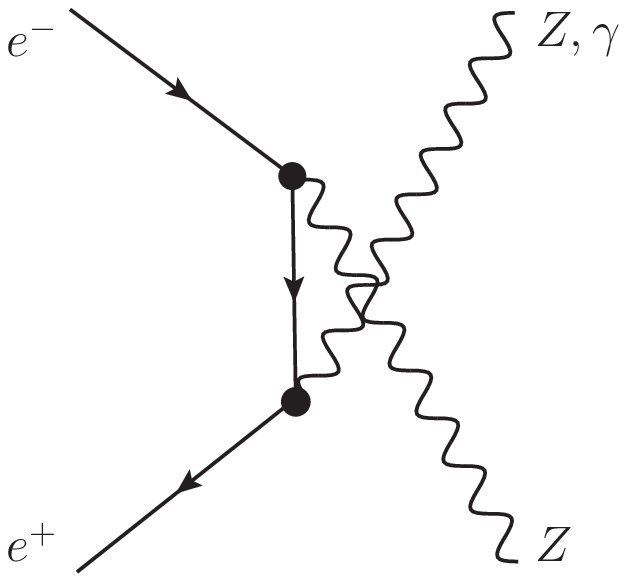}
\hspace{0.5cm}
\includegraphics[width=3.6cm]{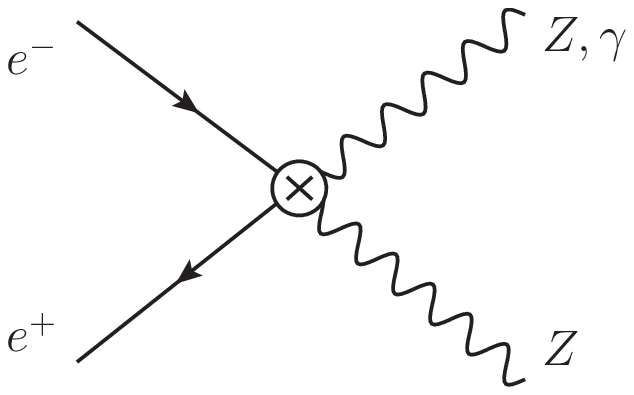}
\\
\vspace{0.2cm}
\includegraphics[width=5.4cm]{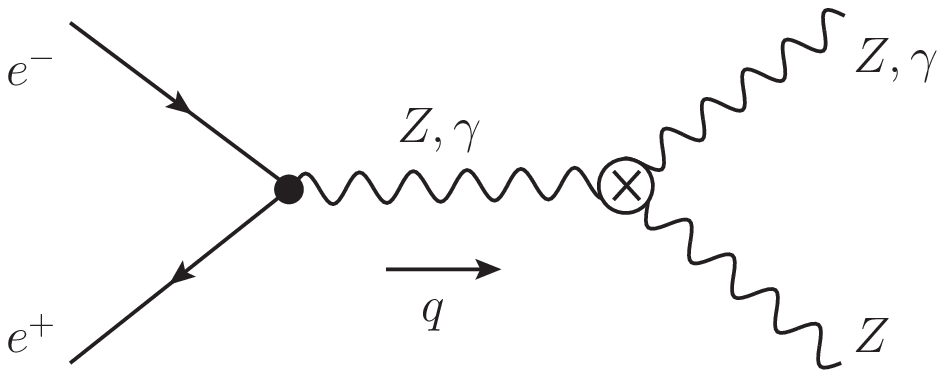}
\end{center}
\caption{\small{\it{Different topologies contributing to neutral gauge boson production. Dotted vertices start at leading order, while crossed ones appear at NNLO.}}}\label{fig:1}
\end{figure*}

In this Section I will discuss all the possible sources of new physics affecting ${\bar{f}}f\to ZZ$ and ${\bar{f}}f\to \gamma Z$. In Fig.~\ref{fig:1} I have listed the different topologies that contribute to neutral gauge boson pair production. The leading contribution comes from the standard model $t$ and $u$-channel exchanges (first two diagrams) while new physics effects generate also $s$-channel and contact term interactions. I will first discuss the structure of the triple gauge boson vertices and later on devote some time to the remaining new physics-induced contributions. 

For generic on-shell gauge bosons (or off-shell but coupled to a conserved current), one can show that the most general expression parametrizing the triple gauge vertex contains just 7 independent kinematical structures~\cite{Hagiwara:1986vm}:  
\begin{align}
\Gamma_{\mu\nu\lambda}&(p_1,p_2;q)=f_1Q_{\lambda}g_{\mu\nu}+f_2(p_{1\nu}g_{\mu\lambda}-p_{2\mu}g_{\nu\lambda})\nonumber\\
&+if_3(p_{1\nu}g_{\mu\lambda}+p_{2\mu}g_{\nu\lambda})+f_4p_{1\nu}p_{2\mu}Q_{\lambda}\nonumber\\
&+if_5\epsilon_{\mu\nu\lambda\rho}Q^{\rho}+f_6\epsilon_{\mu\nu\lambda\rho}q^{\rho}+f_7Q_{\lambda}\epsilon_{\mu\nu\alpha\rho}p_{1}^{\alpha}p_{2}^{\rho}
\end{align}
where $q=p_1+p_2$ and $Q=p_1-p_2$. The previous expression applies, for example, to $WW$ pair production through $Z$-exchange. For $WW$ production through photon exchange a further restriction applies: the structure $(W_{\mu\nu}^+W^{\mu-}-W_{\mu\nu}^-W^{\mu+})A^{\nu}$ is fixed by the charge of the $W$ and thus does not get renormalized. Moreover, gauge invariance implies that $f_{3,5}\sim s$~\cite{Hagiwara:1986vm}. This means that for an on-shell photon such terms are absent. In fact, as I will discuss in more detail in the next Section, such contributions can always be reshuffled through gauge field redefinitions. As a result, and without loss of generality, new physics effects in the $\gamma^*WW$ vertex can be reduced to 4 independent kinematical structures. 

Further reductions occur for the neutral case. For both $\gamma^*ZZ$ and $Z^*ZZ$, Bose symmetry dramatically simplifies the vertex to two structures, namely
\begin{align}\label{ZZ}
\Gamma_{\mu\nu\lambda}^{ZZV}(p_1,p_2;q)&=i\mu_1^V(p_{1\nu}g_{\mu\lambda}+p_{2\mu}g_{\nu\lambda})+i\mu_2^V\epsilon_{\mu\nu\lambda\rho}Q^{\rho}
\end{align}
It can be shown that all the coefficients above scale as $\mu_i^V\sim (s-m_V^2)$. For $\gamma^*ZZ$ this is a consequence of gauge invariance, while for $Z^*ZZ$ it can be eventually traced back to Bose symmetry.

For $\gamma^*\gamma Z$ and $Z^*\gamma Z$, gauge invariance restricts the form factors to take the form
\begin{align}\label{gammaZ}
&\Gamma_{\mu\nu\lambda}^{Z\gamma V}(p_1,p_2;q)=i\beta_1^V\left(p_{2\lambda}g_{\mu\nu}-p_{2\mu}g_{\nu\lambda}\right)+i\beta_2^V\epsilon_{\mu\nu\lambda\rho}p_{2}^{\rho}\nonumber\\
&+i\frac{\beta_3^V}{\Lambda^2}p_{2\mu}(p_1\cdot p_2g_{\nu\lambda}-p_{1\nu}p_{2\lambda})+i\frac{\beta_4^V}{\Lambda^2}Q_{\lambda}\epsilon_{\mu\nu\alpha\beta}p_1^{\alpha}p_2^{\beta}
\end{align}
with $\beta_i^V\sim (s-m_V^2)$. Terms proportional to the Levi-Civit\`a tensor in Eqs.~(\ref{ZZ}) and (\ref{gammaZ}) are CP conserving (but P violating), while the remaining terms violate CP. Since I will be interested in the (leading) linear new physics corrections, which come from the interference of new physics and the standard model, only the CP conserving structures will contribute. Furthermore, it is clear from Eq.~(\ref{gammaZ}) that the last line is parametrically suppressed. In an EFT language, such terms are generated at NNNLO and can be safely neglected. Therefore, only the 4 parameters $\mu_2^V, \beta_2^V$ will be of relevance in the forecoming analysis.  

I will now turn my attention to the remaining new physics contributions. As discussed in the Introduction, the only way of taking into account all possible corrections consistent with gauge symmetry is to work within an effective field theory. As a result of gauge symmetry, the form of the corrections induced by the effective operators will not only affect the vertices of Fig.~\ref{fig:1}, but will also effect shifts both on the gauge boson kinetic terms and the fundamental standard model parameters.  

Quite generically, gauge-fermion interactions will receive new physics contributions in the form 
\begin{align}
{\cal{L}}_f&=e{\bar{\psi}}\gamma_{\mu}A^{\mu}\psi+e\sum_{j=L,R}\bigg[\zeta_j^{(0)}+\delta\zeta_j\bigg]{\bar{\psi}}_j\gamma_{\mu}Z^{\mu}\psi_j
\end{align} 
where $\zeta_L^{(0)}=t_{2W}^{-1}$ and $\zeta_R^{(0)}=-t_{W}$ are the standard model tree level values. 

Gauge boson kinetic term corrections can be parametrized as
\begin{align}
{\cal{L}}_K&=-\frac{1}{4}(1-2\Delta_Z)Z_{\mu\nu}Z^{\mu\nu}-\frac{1}{4}(1-2\Delta_A)A_{\mu\nu}A^{\mu\nu}\nonumber\\
&+\frac{1}{2}\Delta_{AZ}A_{\mu\nu}Z^{\mu\nu}
\end{align}
Finally, since gauge invariance forbids corrections to $\alpha_{em}$, shifts in the standard model fundamental parameters can only affect $m_Z$ and $G_F$ as $m_Z\to (1+\delta_M)m_Z$ and $G_F\to (1+2\delta_G)G_F$.

The best strategy is to reabsorb the kinetic gauge mixing terms and the shifts in the parameters into vertex corrections~\cite{Holdom:1990xq}. Kinetic terms are canonically normalized with the redefinitions:
\begin{align}
Z_{\mu}&=(1+\Delta_Z){\hat{Z}}_{\mu}\nonumber\\
A_{\mu}&=(1+\Delta_A){\hat{A}}_{\mu}+\Delta_{AZ}{\hat{Z}}_{\mu}
\end{align}
Accordingly, the standard model fundamental parameters have to be renormalized as 
\begin{align}
m_Z&=(1-\delta_M-\Delta_Z)\,{\hat{m}}_Z\nonumber\\
G_F&=(1-2\delta_G)\,{\hat{G}}_F\nonumber\\
e&=(1-\Delta_A)\,{\hat{e}}
\end{align}
where $\delta_j$ characterize the direct new physics contributions, while $\Delta_j$ are the shifts due to the gauge boson kinetic normalization. The previous equations imply that
\begin{align}
s_W&=\left(1-\Delta_W\right){\hat{s}}_W
\end{align}
where 
\begin{align}
\Delta_W&=\frac{c_W^2}{c_{2W}}\left(\Delta_A-\Delta_Z-\delta_M-\delta_G\right)
\end{align}
The net effect is therefore a correction to the gauge-fermion couplings as
\begin{align}\label{g-f}
\delta {\hat{\zeta}}_L=\delta \zeta_L-\zeta_L^{(0)}\frac{\delta_M+\delta_G}{c_{2W}^2}+(\Delta_A-\Delta_Z)t_{2W}+\Delta_{AZ}\nonumber\\
\delta {\hat{\zeta}}_R=\delta \zeta_R+\zeta_R^{(0)}\frac{\delta_M+\delta_G}{c_{2W}}+(\Delta_A-\Delta_Z)t_{2W}+\Delta_{AZ}\nonumber\\
\end{align}

Explicit expressions for the different new physics parameters, namely $\mu_2^V$, $\beta_2^V$, $\delta\zeta_{L,R}$, $\delta_{M,G}$ and $\Delta_{A,Z,AZ}$ in terms of EFT coefficients will be given in the next Section.
 

\section{Effective field theory analysis}\label{sec:III}

As discussed in the Introduction, given the present status on the nature of the scalar sector of the standard model, it seems warranted to perform the analysis both in effective field theories with linearly and nonlinearly realized electroweak symmetry breaking. The differences between both realizations are of profound dynamical significance and affect the nature of the new physics scale: in the linear theory, the new physics scale $\Lambda$ decouples and dimensional power-counting applies. In contrast, in the nonlinear theory the scale of new physics is dynamically generated and therefore bound to be $\Lambda\sim 4\pi v\sim 3$ TeV. This invalidates the naive dimensional power-counting, which has to be substituted by a more elaborate one~\cite{Buchalla:2012qq}. Despite the differences, for processes like gauge boson pair production, in which the scalar sector is not directly involved, both effective field theories should differ only slightly. In $WW$ production, for instance, the number of operators on both theories is roughly the same, but in the linear case some of them are further suppressed ({\it{i.e.}}, they are shifted from NLO to NNLO)~\cite{Buchalla:2013wpa}. 

In the following I will be working with the general nonlinearly-realized effective field theory developed in Ref.~\cite{Buchalla:2012qq} and only later I will compare with the linear realization. In the (minimal) nonlinear framework, electroweak symmetry breaking is realized by spontaneously breaking a global $SU(2)_L\times SU(2)_R$ down to $SU(2)_V$. The resulting Goldstone modes are then collected into a matrix $U$ transforming as $g_L U g_R^{\dagger}$ under the global group. One also defines
\begin{align}
D_\mu U=\partial_\mu U+i g W_\mu U -i g' B_\mu U T_3
\end{align}
such that the standard model subgroup $SU(2)_L\times U(1)_Y$ is gauged. Henceforth I will use the shorthand notation \begin{align}
L_{\mu}=iUD_{\mu}U^{\dagger};\qquad \tau_L=UT_3U^{\dagger}
\end{align}
for the Goldstone covariant derivative and the custodial symmetry breaking spurion $T_3$. With these definitions, at leading order one gets
\begin{align}
{\cal{L}}_{LO}&=-\frac{1}{2}\langle W_{\mu\nu}W^{\mu\nu}\rangle 
-\frac{1}{4} B_{\mu\nu}B^{\mu\nu}+\frac{v^2}{4}\ \langle L_\mu L^\mu\rangle\nonumber\\
&+i\sum_{j=L,R}\bar {\psi}_j \!\not\!\! D\psi_j+{\cal{L}}_{Yukawa}(U,{\bar{f}}_j,f_j)
\end{align}   
where 
\begin{align}
D_{\mu}\psi_L&=(\partial_{\mu}+igW_{\mu}+ig^{\prime}Y_LB_{\mu})\psi_L\nonumber\\ D_{\mu}\psi_R&=(\partial_{\mu}+ig^{\prime}Y_RB_{\mu})\psi_R
\end{align}  
In this work I will be only concerned with first-family leptons, such that the Yukawa terms will be negligible.

NLO operators can be generically defined as
\begin{align}
{\cal{L}}_{NLO}&=\beta {\cal{O}}_{\beta}+\sum_{j}c_j\frac{v^{6-d_j}}{\Lambda^2}{\cal{O}}_j
\end{align}
where $d_j$ is the spacetime dimension of ${\cal{O}}_j$. From the previous equation it is clear that due to the nondecoupling nature of the interactions, naive dimensional power-counting is not at work. NLO operators were fully classified and worked out in~\cite{Buchalla:2012qq}. With $U,\psi,X$ as shorthand notations for Goldstone, fermion and gauge fields, the classes were denoted as $UD^4$, $XUD^2$, $X^2U$, $\psi^2UD$, $\psi^2UD^2$ and $\psi^4U$. Out of this, the only relevant classes for ${\bar{f}}f\to \gamma Z,ZZ$ at NLO are $X^2U$, $\psi^2UD$ and $\psi^4U$. The relevant subset of operators is
\begin{align}
{\cal{L}}_{NLO}&=\sum_j^2 \lambda_j {\cal{O}}_{Xj}+\sum_j \eta_j {\cal{O}}_{Vj}+\beta{\cal{O}}_{\beta}+\eta_{4f}{\cal{O}}_{4f}
\end{align}
where
\begin{align}\label{naive1}
{\cal{O}}_{X1}&=g^{\prime}gB_{\mu\nu}\langle W^{\mu\nu}\tau_L\rangle\nonumber\\
{\cal{O}}_{\beta}&=v^2\langle \tau_LL_{\mu}\rangle^2\nonumber\\
{\cal{O}}_{X2}&=g^2 \langle W^{\mu\nu}\tau_L\rangle^2
\end{align}
are oblique corrections,
\begin{align}
{\cal O}_{V1}&=-\bar q\gamma^\mu q\ \langle L_{\mu}\tau_L\rangle ,\,\, &&
{\cal O}_{V4}=-\bar u\gamma^\mu u\ \langle L_{\mu}\tau_L\rangle \nonumber\\
{\cal O}_{V2}&=-\bar q\gamma^\mu \tau_L q\ 
\langle L_{\mu}\tau_L\rangle , \,\, &&
{\cal O}_{V5}=-\bar d\gamma^\mu d\ \langle L_{\mu}\tau_L\rangle \nonumber\\
{\cal O}_{V7}&=-\bar l\gamma^\mu l\ \langle L_{\mu}\tau_L\rangle ,\,\, &&
{\cal O}_{V9}=-\bar l\gamma^\mu \tau_{12} l\ 
\langle  L_{\mu}\tau_{21}\rangle 
\nonumber\\
{\cal O}_{V8}&=-\bar l\gamma^\mu \tau_L l\ 
\langle L_{\mu}\tau_L\rangle \,\, &&
{\cal O}_{V10}=-\bar e\gamma^\mu e\ \langle L_{\mu}\tau_L\rangle
\end{align}
are gauge-fermion new physics contributions ($\tau_{12}=T_1+iT_2$; $\tau_{21}=T_1-iT_2$), and
\begin{align}
{\cal{O}}_{4f}&=\frac{1}{2}({\cal{O}}_{LL5}-4{\cal{O}}_{LL15})=({\bar{e}}_L\gamma_{\rho}\mu_L)({\bar{\nu}}_{\mu}\gamma^{\rho}\nu_e)
\end{align}
${\cal{O}}_{X1,2}$, ${\cal{O}}_{\beta}$, ${\cal{O}}_{V9}$ and ${\cal{O}}_{4j}$ are indirect contributions: the first two shift the kinetic terms, the third renormalizes $m_Z$ and the remaining two renormalize $G_F$. ${\cal{O}}_{V1-5}$ are the 4 direct contributions to gauge-fermion vertices in $pp$ collisions (correcting left and right-handed ${\bar{u}}uZ$ and ${\bar{d}}dZ$ vertices), while for $e^+e^-$ ${\cal{O}}_{V10}$ and the combination $\frac{1}{2}{\cal{O}}_{V7}-{\cal{O}}_{V8}$ respectively correct the right and left-handed $e^+e^-Z$ vertices.    

Comparing with the phenomenological parameters introduced in the previous Section, one finds the following matching equations:
\begin{align}
\delta_M&=-\beta\nonumber\\
\delta_G&=\eta_9-\frac{v^2}{4\Lambda^2}\eta_{4f}\nonumber\\
\Delta_Z&=-\lambda_1+\frac{\lambda_2}{2 t_W^2}\nonumber\\
\Delta_A&=\lambda_1+\frac{\lambda_2}{2}\nonumber\\
\Delta_{AZ}&=\frac{2\lambda_1}{t_{2W}}+\frac{\lambda_2}{t_W}
\end{align}
which are generic to both $pp$ and $e^+e^-$ collisions. The previous set of equations, once plugged into Eqs.~(\ref{g-f}), lead to:
\begin{align}\label{gfEFT}
\delta {\hat{\zeta}}_L&=\frac{2e^2\lambda_1-c_{2W}\eta_L+{\hat{\beta}}}{s_{2W}c_{2W}}\nonumber\\
\delta {\hat{\zeta}}_R&=\frac{2e^2\lambda_1-c_{2W}\eta_R+2s_W^2{\hat{\beta}}}{s_{2W}c_{2W}}
\end{align}
where we have defined ${\hat{\beta}}=-(\delta_M+\delta_G)\simeq \beta-\eta_9$ and the parametrically-suppressed $\eta_{4f}$ contribution has been neglected. $\{\eta_L=\eta_7-\eta_8/2;\,\eta_R=\eta_{10}\}$ for $e^+e^-$ collisions and $\{\eta_L=f_u(\eta_7+\eta_8/2)+f_d(\eta_7-\eta_8/2);\,\eta_R=g_u(\eta_{4})+g_d(\eta_5)\}$ for $pp$ collisions, where $f_{u,d}$ and $g_{u,d}$ are a reminder that the coefficients have to be dressed with the appropriate parton distribution functions.

It is interesting to point out that in the case of $e^+e^-$ collisions the gauge-boson corrections in Eqs.~(\ref{gfEFT}) can be cast entirely in terms of the oblique $S$ and $T$ parameters. In general, the relevant 3 gauge-fermion operators can be eliminated in favor of 3 charged triple gauge coefficients and the oblique parameters~\cite{Buchalla:2012qq,Buchalla:2013wpa}. However, since charged triple gauge operators do not contribute to the neutral processes, in practice this means that one can eliminate the gauge-fermion operators and express the gauge-fermion corrections of Eqs.~(\ref{gfEFT}) entirely in terms of (redefined) $\lambda_1$ and $\beta$ coefficients, which can be related to the oblique parameters as $S=-16\pi\lambda_1$ and $\alpha T=2\beta$. Thus one finds
\begin{align}\label{ST}
\delta\zeta_L&=\frac{\alpha}{s_{4W}}(T-S)\nonumber\\
\delta\zeta_R&=\frac{\alpha}{s_{4W}}(2s_W^2T-S)
\end{align}
I want to emphasize that the manipulations leading to Eqs.~(\ref{ST}) are only valid for $e^+e^-$ collisions (and in the nonlinear basis). In $pp$ collisions, the increased number of gauge-fermion operators does not allow to remove them altogether.\footnote{Notice that even though some of the gauge-fermion coefficients in $pp$ collisions can be replaced by $\lambda_1$ and $\beta$, one can no longer relate them to $S$ and $T$, which are defined for vanishing gauge-fermion operators.}

The most relevant difference between charged and neutral gauge boson pair production is that in the former the contribution to the triple gauge vertex arises at NLO with direct contributions from the ${\cal{O}}_{X1-6}$ operators. In the neutral case, there are no direct contributions at NLO. This can be seen from the fact that $\mu_i^V,\beta_i^V\sim (s-m_V^2)$, which implies that contributions to neutral triple gauge vertices have to come at least at NNLO. At that order the effective Lagrangian generically reads 
\begin{align}
{\cal{L}}_{NNLO}&=\sum_{j}c_j\frac{v^{8-d_j}}{\Lambda^4}{\cal{O}}_j
\end{align} 
In Ref.~\cite{Buchalla:2012qq} the NNLO classes were identified to be $UD^6$, $UXD^4$, $X^2UD^2$, $\psi^2 UD^3$, $X^3U$, $\psi^2 UX$ and $\psi^2 UXD$ but the corresponding operators were not classified. If one works in the unitary gauge, contributions to ${\bar{f}}f\to ZZ,\gamma Z$ can potentially come only from $X^3U$, $\psi^2 UX$ and $\psi^2UXD$, all the other classes having far too many gauge bosons. $X^3U$ and $\psi^2 UX$ produce corrections to charged triple gauge vertices, thus leaving $\psi^2UXD$ as the only potential class for neutral triple gauge vertices. Without much effort, one can show that the relevant operators inside this class are
\begin{align}\label{basis}
&{\cal{L}}_{nTGV}\!=\!\!\!\!\sum_{j=L,R}\!\!\bigg\{\frac{c_{jW}}{\Lambda^2}J_{\mu}^{(j)}\langle W^{\mu\nu}L_{\nu} \rangle+\frac{c_{jB}}{\Lambda^2}J_{\mu}^{(j)} B^{\mu\nu} \langle \tau_L L_{\nu} \rangle\nonumber\\
&+\frac{{\tilde{c}}_{jW}}{\Lambda^2}J_{\mu}^{(j)}\langle {\tilde{W}}^{\mu\nu}L_{\nu} \rangle+\frac{{\tilde{c}}_{jB}}{\Lambda^2}J_{\mu}^{(j)} {\tilde{B}}^{\mu\nu} \langle \tau_L L_{\nu} \rangle\bigg\}
\end{align}
where $J_{\mu}^{(L,R)}=({\bar{f}}\gamma_{\mu}f)_{L,R}$. 

Notice that the previous operators correspond to the contact term diagram in Fig.~\ref{fig:1}, while there is no contribution to the $s$-channel. This might sound surprising, since after all we are after the effects of triple gauge vertices. However, this is perfectly consistent: since neutral triple gauge vertices are proportional to $(s-m_V^2)$, their effects can always be reabsorbed in the form of contact terms. In other words, the effective operators that can be constructed to affect nTGC can always be reduced using the equations of motion to operators with fermion fields~\cite{Gounaris:1999kf}.  Thus, the $s$-channel and contact configurations in Fig.~\ref{fig:1} are not independent but complementary, depending on the choice of effective operator basis. 
  
In order to see explicitly this equivalence and make contact with the triple gauge vertex coefficients, use has to be made of the equations of motion for the gauge fields: 
\begin{align}
\partial^{\mu}B_{\mu\nu}&=g^{\prime}\left[\sum_jY_j{\bar{f}}_j\gamma_{\nu}f_j+\frac{v^2}{2}\langle \tau_LL_{\nu}\rangle\right];\nonumber\\
D^{\mu}W_{\mu\nu}^a&=g\left[\sum_j{\bar{f}}_{jL}\gamma_{\nu}T^a f_{jL}-\frac{v^2}{2}\langle T^aL_{\nu}\rangle\right]
\end{align}
At linear colliders, one can restrict the fermion bilinears above to $e^+e^-$. In the unitary gauge one then finds that
\begin{align}\label{basischange}
{\bar{l}}\gamma_{\nu}l&=-\frac{2s_W}{e}\Big[c_W \partial^{\mu}Z_{\mu\nu}+s_W \partial^{\mu}F_{\mu\nu}\Big]\nonumber\\
{\bar{e}}\gamma_{\nu}e&=\frac{1}{e}\Big[s_{2W}\partial^{\mu}Z_{\mu\nu}-c_{2W}\partial^{\mu}F_{\mu\nu}\Big]
\end{align}
and, using that (also in the unitary gauge)
\begin{align}
\langle W_{\mu\nu}L^{\nu} \rangle &=\frac{e}{s_{2W}}Z^{\nu}W_{\mu\nu}^{(3)}+\frac{e}{2s_W}\sum_{a=1,2}W^{\nu(a)}W_{\mu\nu}^{(a)}\nonumber\\
\langle \tau_L L_{\nu} \rangle &=\frac{e}{s_{2W}}Z_{\nu}
\end{align}
Eq.~(\ref{basis}) can be rewritten as
\begin{align}
{\cal{L}}_{nTGV}&=\frac{\lambda_{ZZ}}{\Lambda^2}\partial_{\lambda}Z^{\lambda\mu}Z^{\nu}Z_{\mu\nu}+\frac{\lambda_{\gamma\gamma}}{\Lambda^2}\partial_{\lambda}F^{\lambda\mu}Z^{\nu}F_{\mu\nu}\nonumber\\
&+\frac{\lambda_{Z\gamma}}{\Lambda^2}\partial_{\lambda}Z^{\lambda\mu}Z^{\nu}F_{\mu\nu}+\frac{\lambda_{\gamma Z}}{\Lambda^2}\partial_{\lambda}F^{\lambda\mu}Z^{\nu}Z_{\mu\nu}\nonumber\\
&+\frac{{\tilde{\lambda}}_{ZZ}}{\Lambda^2}\partial_{\lambda}Z^{\lambda\mu}Z^{\nu}{\tilde{Z}}_{\mu\nu}+\frac{{\tilde{\lambda}}_{\gamma\gamma}}{\Lambda^2}\partial_{\lambda}F^{\lambda\mu}Z^{\nu}{\tilde{F}}_{\mu\nu}\nonumber\\
&+\frac{{\tilde{\lambda}}_{Z\gamma}}{\Lambda^2}\partial_{\lambda}Z^{\lambda\mu}Z^{\nu}{\tilde{F}}_{\mu\nu}+\frac{{\tilde{\lambda}}_{\gamma Z}}{\Lambda^2}\partial_{\lambda}F^{\lambda\mu}Z^{\nu}{\tilde{Z}}_{\mu\nu}
\end{align}
which is in agreement with Ref.~\cite{Gounaris:2000dn,Alcaraz:2001nv}. Both basis of operators can be related by 
\begin{align}\label{ident}
\lambda_{ZZ}&=-c_Wc_{LW}+s_Wc_{LB}+c_Wc_{RW}-s_Wc_{RB}\nonumber\\
\lambda_{\gamma\gamma}&=-\frac{s_W^2}{c_W}c_{LW}-s_Wc_{LB}-\frac{s_W}{t_{2W}}c_{RW}-\frac{c_W}{t_{2W}}c_{RB}\nonumber\\
\lambda_{Z\gamma}&=-s_Wc_{LW}-c_Wc_{LB}+s_Wc_{RW}+c_Wc_{RB}\nonumber\\
\lambda_{\gamma Z}&=-s_Wc_{LW}+\frac{s_W^2}{c_W}c_{LB}-\frac{c_W}{t_{2W}}c_{RW}+\frac{s_W}{t_{2W}}c_{RB}
\end{align} 
with analogous expressions holding for the dual operators. Using this basis, nTGC for $ZZ$ and $\gamma Z$ production can be expressed as
\begin{equation}\label{ident1}
\begin{array}{ll}
\mu_1^Z=\lambda_{ZZ}\displaystyle\frac{s-m_Z^2}{\Lambda^2};\qquad &
\mu_1^{\gamma}=\lambda_{\gamma Z}\displaystyle\frac{s}{\Lambda^2}\\
\mu_2^Z=-{\tilde{\lambda}}_{ZZ}\displaystyle\frac{s-m_Z^2}{\Lambda^2};\qquad &
\mu_2^{\gamma}=-{\tilde{\lambda}}_{\gamma Z}\displaystyle\frac{s}{\Lambda^2}
\end{array}
\end{equation} 
and
\begin{equation}\label{ident2}
\begin{array}{ll}
\beta_1^Z=-\lambda_{Z\gamma}\displaystyle\frac{s-m_Z^2}{\Lambda^2};\qquad &
\beta_1^{\gamma}=-\lambda_{\gamma\gamma}\displaystyle\frac{s}{\Lambda^2}\\
\beta_2^Z={\tilde{\lambda}}_{Z\gamma}\displaystyle\frac{s-m_Z^2}{\Lambda^2};\qquad &
\beta_2^{\gamma}={\tilde{\lambda}}_{\gamma\gamma}\displaystyle\frac{s}{\Lambda^2}
\end{array}
\end{equation}   
which indeed have the correct $\mu_j^V,\beta_j^V\sim (s-m_V^2)$ behavior. Intuitively, the third and fourth diagrams in Fig.~\ref{fig:1} are equivalent because the structure of the triple gauge coefficients cancels the $\gamma$ and $Z$ propagators, effectively shrinking the contributions to a contact term. 

However, it is worth emphasizing that trading the contact terms for the $s$-channel contributions is exclusive from $e^+e^-$ collisions. Eqs.~(\ref{basischange}) only hold for this particular process. In general, for instance for $pp$ collisions, where 4 gauge-fermion currents are needed, one can only eliminate the contact terms partially. As a result, in a full-fledged EFT analysis of new physics in neutral gauge boson pair production at hadron colliders one cannot eliminate the contact term in Fig.~\ref{fig:1} but instead can eliminate the $s$-channel contribution. This is so because there are more operators involving fermion bilinears than gauge boson self-interactions.

So far my discussion has concentrated on the nonlinearly-realized effective field theory of the electroweak interactions. Since no scalar particles are involved in the analysis, one expects very little deviations in the linear basis, except for occasional shiftings of the operators to higher dimensionality, as happens in $WW$ production (see the discussion in Ref.~\cite{Buchalla:2013wpa}). In the neutral case, it so turns out that every NLO operator discussed above is in a one-to-one correspondence with a corresponding operator in the linearly-realized EFT without any dimensionality shifts whatsoever: notice that the only potential exception, namely ${\cal{O}}_{X2}$ (NNLO in the linear basis) is eventually irrelevant in the analysis.

Regarding the NNLO contact term operators in Eq.~(\ref{basis}), the linear ones can be straightforwardly constructed by the following replacements: 
\begin{align}
\langle L_{\mu}\tau_L\rangle&\to\frac{i}{2}(\phi^{\dagger}\stackrel{\leftrightarrow}{D}_{\mu}\phi)\nonumber\\
\langle W^{\mu\nu}L_{\nu}\rangle&\to-\frac{i}{4}W^{\mu\nu(a)}(\phi^{\dagger}\stackrel{\leftrightarrow}{D_{\nu}^a}\phi)
\end{align}
where $\stackrel{\leftrightarrow}{D}_{\mu}=\stackrel{\rightarrow}{D}_{\mu}-\stackrel{\leftarrow}{D}_{\mu}$ and  $\stackrel{\leftrightarrow}{D_{\nu}^a}=T^a\stackrel{\rightarrow}{D}_{\mu}-\stackrel{\leftarrow}{D}_{\mu}T^a$. 
Therefore, in the linear basis there are also 8 independent NNLO operators contributing to the triple gauge vertex. This agrees with the conclusions of Ref.~\cite{Alcaraz:2001nv} and contrasts with the results obtained in~\cite{Larios:2000ni} and the statement made in~\cite{Gounaris:2000dn} that there are just 2 independent operators. This latter statement, which is still used in some experimental analyses to constraint the new physics coefficients, is incorrect. 
 

\section{Polarized cross sections at high energies}\label{sec:IV}
In this Section I will present the results for $e^+e^-\to \gamma Z$ and $e^+e^-\to ZZ$ for different initial and final-state polarizations. I will show that gauge-fermion and triple gauge corrections leave rather distinct signatures. While the former appear as tiny corrections to the dominant standard model background in purely transversal (TT) production, the latter are parametrically dominant in LT-polarized final states.

To fix my conventions, $Z_{\mu}^{(1)}$ will be chosen to point in the positive $z$ direction, and $\theta$ will be the angle between the incoming $e^-$ and $Z_{\mu}^{(1)}$. With these conventions,
\begin{align}
q_1^{\mu}&=\frac{\sqrt{s}}{2}\left(\begin{array}{c}1\\-\sin\theta\\0\\ \cos\theta\end{array}\right);\qquad q_2^{\mu}=\frac{\sqrt{s}}{2}\left(\begin{array}{c}1\\ \sin\theta\\0\\ -\cos\theta\end{array}\right);\nonumber\\
\epsilon_{+1}^{(1)}&=\epsilon_{-1}^{(2)}=\frac{1}{\sqrt{2}}\left(\begin{array}{c}0\\-1\\-i\\ 0\end{array}\right);\quad
\epsilon_{-1}^{(1)}=\epsilon_{+1}^{(2)}=\frac{1}{\sqrt{2}}\left(\begin{array}{c}0\\1\\-i\\ 0\end{array}\right)
\end{align}
For $ZZ$ production one has
\begin{align}
p_1^{\mu}&=\frac{\sqrt{s}}{2}\left(\begin{array}{c}1\\0\\0\\ \beta \end{array}\right);\qquad
p_2^{\mu}=\frac{\sqrt{s}}{2}\left(\begin{array}{c}1\\0\\0\\ -\beta\end{array}\right);\nonumber\\
\epsilon_{0}^{(1)}&=\frac{\sqrt{s}}{2m_Z}\left(\begin{array}{c}\beta\\0\\0\\ 1\end{array}\right);\qquad \epsilon_{0}^{(2)}=\frac{\sqrt{s}}{2m_Z}\left(\begin{array}{c}-\beta\\0\\0\\ 1\end{array}\right)
\end{align}
with $\beta=\sqrt{1-\frac{4m_Z^2}{s}}$, while for $\gamma Z$ production
\begin{align}
p_1^{\mu}&=\frac{\sqrt{s}}{2}\left(\begin{array}{c}b_+\\0\\0\\ b_- \end{array}\right);\qquad
p_2^{\mu}=\frac{\sqrt{s}}{2}\left(\begin{array}{c}b_-\\0\\0\\ -b_-\end{array}\right);\nonumber\\
\epsilon_{0}^{(1)}&=\frac{\sqrt{s}}{2m_Z}\left(\begin{array}{c}b_-\\0\\0\\ b_+\end{array}\right);\qquad b_{\pm}=1\pm\frac{m_Z^2}{s}
\end{align}
In the following I will discuss each process separately.

\subsection{$e^+e^-\to \gamma Z$}
The interference of standard model and new physics effects results in:
\begin{align}
\frac{d\sigma_{\gamma Z}^j}{dt}&=\frac{e^4\zeta_j^2(m_Z^4+s^2-2tu)}{4\pi s^2tu}\nonumber\\
&+\frac{e^3\zeta_j(m_Z^2+s)}{4\pi s^2}\left[\frac{\beta_{2A}}{s}+\frac{\zeta_j\beta_{2Z}}{s-m_Z^2}\right]
\end{align}
where $j=L(R)$ denotes the contributions from purely left-handed (right-handed) polarized electrons. $\zeta_j=\zeta_j^{(0)}+\delta\zeta_j$, where $\zeta_L^{(0)}=t_{2W}^{-1}$, $\zeta_R^{(0)}=-t_W$ and $\delta\zeta_j$ are given in Eqs.~(\ref{gfEFT}). The nTGV parameters $\beta_{2A,2Z}$ can be expressed in terms of ${\tilde{\lambda}}_{Z\gamma},{\tilde{\lambda}}_{\gamma\gamma}$. However, it is more instructive to express them in terms of ${\tilde{c}}_{jW},{\tilde{c}}_{jB}$:  
\begin{align}\label{comb}
\frac{\beta_{2A}}{s}+\frac{\zeta_j\beta_{2Z}}{s-m_Z^2}&=-\frac{1}{2\Lambda^2}\left[\frac{{\tilde{c}}_{jW}}{c_W}+\frac{{\tilde{c}}_{jB}}{s_W}\right]
\end{align}  
from which it is clear that initial-state polarizations match.

At a linear collider, the hierarchy $v\ll\sqrt{s}\ll \Lambda$ holds, and one can expand the results in powers of $v^2/s$. The tree level standard model result is then
\begin{align}
\frac{d\sigma_{SM}^j}{d\cos{\theta}}&=\frac{4\pi\alpha^2}{s}\left(\frac{1+\cos^2{\theta}}{\sin^2{\theta}}\right)\zeta_j^2
\end{align}
while new physics corrections take the form:
\begin{align}
\frac{d\sigma_{BSM}^j}{d\cos{\theta}}&=\frac{8\pi\alpha^2}{s}\zeta_j\left[\left(\frac{1+\cos^2{\theta}}{\sin^2{\theta}}\right)\delta\zeta_j\right.\nonumber\\
&\left.\,\,\,\,\,\,\,\,\,\,\,\,-\frac{s}{\Lambda^2}\frac{s_W{\tilde{c}}_{jW}+c_W{\tilde{c}}_{jB}}{4es_{2W}}\right]
\end{align}
As mentioned at the beginning of this Section, the two contributions above have very distinct kinematical behavior. This motivates to split the previous result into the different final-state polarizations. The itemized contributions for generic energies are
\begin{align}
\frac{d\sigma_{\gamma Z_L}^j}{dt}&=\frac{e^4m_Z^2\zeta_j^2}{\pi s(m_Z^2-s)^2}+\frac{e^3\zeta_j}{4\pi s}\left[\frac{\beta_{2A}}{s}+\frac{\zeta_j\beta_{2Z}}{s-m_Z^2}\right]\nonumber\\
\frac{d\sigma_{\gamma Z_T}^j}{dt}&=\frac{e^4\zeta_j^2(m_Z^4+s^2)((m_Z^2-s)^2-2tu)}{4\pi s^2tu(m_Z^2-s)^2}\nonumber\\
&+\frac{e^3\zeta_jm_Z^2}{4\pi s^2}\left[\frac{\beta_{2A}}{s}+\frac{\zeta_j\beta_{2Z}}{s-m_Z^2}\right]
\end{align}
In the high energy limit they reduce to
\begin{align}
\frac{d\sigma_{\gamma^{\pm} Z_L}^j}{d\cos\theta}&=-\frac{\pi\alpha^2(1\mp \cos{\theta})\zeta_j}{\Lambda^2} \frac{s_W{\tilde{c}}_{jW}+c_W{\tilde{c}}_{jB}}{es_{2W}}\nonumber\\
\frac{d\sigma_{\gamma^{\mp} Z_{T}^{\pm}}^j}{d\cos\theta}&=\frac{2\pi\alpha^2}{s}\frac{1\pm \cos{\theta}}{1\mp \cos{\theta}}\zeta_j^2
\end{align}
which are valid up to ${\cal{O}}(v^2/s^2)$ corrections. 


\subsection{$e^+e^-\to ZZ$}
One can repeat the steps of the previous subsection for $ZZ$ production. The interference of the standard model and new physics gives
\begin{widetext}
\begin{align}
\frac{d\sigma_{ZZ}^j}{dt}&=-e^4 \zeta_j^4 \frac{4 m_Z^8-4 m_Z^6 (s+3 t)+m_Z^4 (s^2+6 s t+14 t^2)-2 m_Z^2 t (s+2t)^2+t(s^3+3s^2t+4st^2+2t^3)}{4\pi s^2t^2u^2}\nonumber\\
&+e^3\zeta_j^2 \frac{2m_Z^6-m_Z^4 (3s+4t)+m_Z^2 (s^2+2 t^2)+s t (s+t)}{2\pi s^2t u}\left[\frac{\mu_{2A}}{s}+\frac{\zeta_j\mu_{2Z}}{s-m_Z^2}\right]
\end{align}
\end{widetext}
Triple gauge parameters can be likewise expressed in terms of ${\tilde{c}}_{jW}$ and ${\tilde{c}}_{jB}$ coefficients. The result
\begin{align}\label{comb1}
\frac{\mu_{2A}}{s}+\frac{\zeta_j\mu_{2Z}}{s-m_Z^2}&=\frac{1}{2\Lambda^2}\left[\frac{{\tilde{c}}_{jW}}{s_W}-\frac{{\tilde{c}}_{jB}}{c_W}\right]
\end{align} 
shows again that the initial polarizations match. However, since the specific combination of ${\tilde{c}}_{jW}$ and ${\tilde{c}}_{jB}$ differs in $ZZ$ and $\gamma Z$ (compare Eq.~(\ref{comb}) and (\ref{comb1})), in a combined fit one can separately extract ${\tilde{c}}_{jW}$ and ${\tilde{c}}_{jB}$ for each initial-state polarization. 

At high energies the standard model result becomes
\begin{align}
\frac{d\sigma_{SM}^j}{d\cos{\theta}}&=\frac{4\pi\alpha^2}{s}\left(\frac{1+\cos^2{\theta}}{\sin^2{\theta}}\right)\zeta_j^4
\end{align}
and the new physics corrections are
\begin{align}
\frac{d\sigma_{BSM}^j}{d\cos{\theta}}&=\frac{16\pi\alpha^2}{s}\zeta_j^3\left[\left(\frac{1+\cos^2{\theta}}{\sin^2{\theta}}\right)\delta\zeta_j\right.\nonumber\\
&\left.\,\,\,\,\,\,\,\,\,\,\,\,\,\,\,-\frac{s}{\Lambda^2}\frac{c_W{\tilde{c}}_{jW}-s_W{\tilde{c}}_{jB}}{4e\zeta_js_{2W}}\right]
\end{align}
Again, different final-state polarizations can be used to disentangle both contributions above. The results for generic energies for the different channels read
\begin{align}
\frac{d\sigma_{Z_LZ_L}^{j}}{d\cos\theta}&=\frac{8\pi\alpha^2 s m_Z^4\beta\sin^2\!2\theta}{(s^2\beta^2\sin^2\!\theta+4m_Z^4)^2}\zeta_j^4 \nonumber\\
\frac{d\sigma_{Z_{L}Z_{T}^{\pm}}^j}{d\cos\theta}&\!\!=\!\!\frac{4\pi\alpha^2(1\!\pm \cos{\theta})^2 m_Z^2 \beta((1\!\mp \cos{\theta})s-2m_Z^2)^2}{(s^2\beta^2\sin^2\!\theta+4 m_Z^4)^2}\zeta_j^4\nonumber\\
&\!\!\!\!\!\!\!\!\!\!\!\!\!\!\!\!\!\!\!\!\!\!\!\!+\frac{\pi\alpha^2(1\!\pm \cos{\theta})^2\beta^3s((1\!\mp \cos{\theta})s-2m_Z^2)}{s^2\beta^2\cos^2\!\theta-(s-2m_Z^2)^2}\zeta_j^2\nonumber\\
&\times\frac{1}{e}\left[\frac{\mu_{2A}}{s}+\frac{\zeta_j\mu_{2Z}}{s-m_Z^2}\right]\nonumber\\
\frac{d\sigma_{Z_{T}^{\pm}Z_{T}^{\mp}}^{j}}{d\cos\theta}
&=\frac{2\pi\alpha^2 s^2_{\theta}(1\pm \cos\theta)^2\beta s(s-2m_Z^2)^2}{(s^2\beta^2\sin^2\!\theta+4m_Z^4)^2}\zeta_j^4\nonumber\\
\frac{d\sigma_{Z_{T}^{\pm}Z_{T}^{\pm}}^{j}}{d\cos\theta}&=\frac{1}{4}\frac{d\sigma^{LL}}{d\cos\theta}
\end{align}
whose high energy behavior is 
\begin{align}
\frac{d\sigma_{Z_{L}Z_{T}^{\pm}}^j}{d\cos{\theta}}&=-\frac{\pi\alpha^2(1\mp\cos{\theta})\zeta_j^2}{\Lambda^2} \frac{c_W{\tilde{c}}_{jW}-s_W{\tilde{c}}_{jB}}{es_{2W}}\nonumber\\
\frac{d\sigma_{Z_{T}^{\pm}Z_{T}^{\mp}}^j}{d\cos{\theta}}&=\frac{2\pi\alpha^2}{s}\frac{1\pm\cos\theta}{1\mp\cos\theta}\zeta_j^4
\end{align}
The results above are valid up to ${\cal{O}}(v^2/s^2)$ corrections. The LL channel only scales as ${\cal{O}}(v^4/s^3)$ and can be safely neglected. 

\subsection{Numerical analysis}

\begin{figure*}[t]
\begin{center}
\includegraphics[width=7cm]{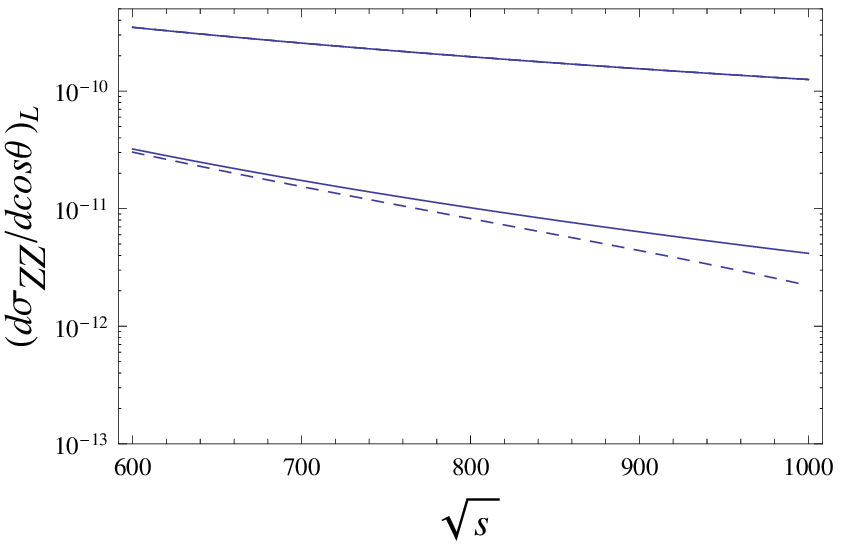}
\hspace{1.0cm}
\includegraphics[width=7cm]{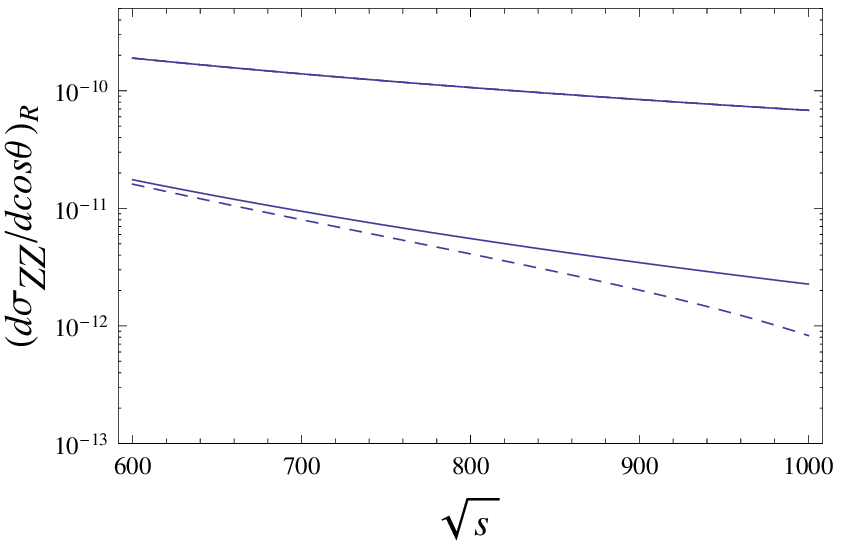}\\
\includegraphics[width=7cm]{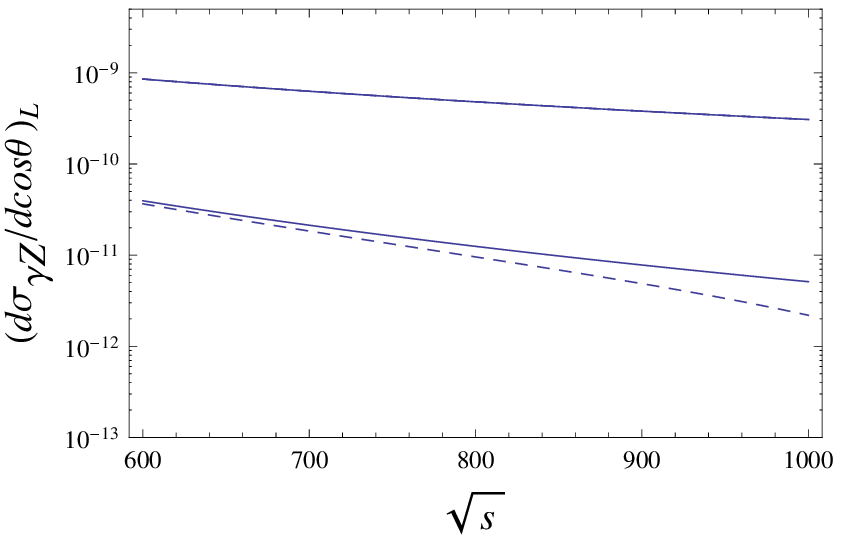}
\hspace{1.0cm}
\includegraphics[width=7cm]{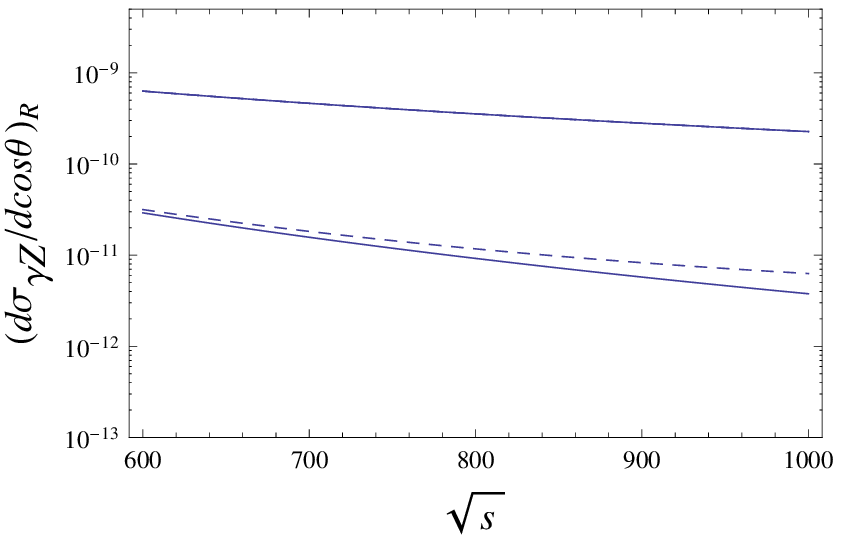}
\end{center}
\caption{\small{\it{TT (solid upper line) and LT (solid lower line) final-state polarizations in the standard model as a function of energy (in GeV) for left-handed and right-handed initial-state polarizations, with $\cos\theta=0$. Dashed lines indicate the typical new physics corrections: for LT it is parametrically enhanced with energy, while for TT it is only a permille correction.}}}\label{fig:2}
\end{figure*}

The results for the different initial and final-state polarizations are shown in Fig.~\ref{fig:2} as a function of energy for $\cos\theta=0$. For simplicity, the analysis has been restricted to $ZZ$ and $\gamma Z$ production in $e^+e^-$ collisions. 
For the standard model parameters I have used
\begin{align}
m_Z=91.19\,{\mathrm{GeV}};\quad s_W^2=0.231;\quad \alpha=1/129
\end{align}
In order to estimate the expected size of the NLO new physics effects, I have used the gauge-fermion corrections in the form of Eq.~(\ref{ST}). The latest Gfitter results yield~\cite{Baak:2011ze}
\begin{align}
S=0.03\pm 0.10;\qquad T=0.05\pm0.12
\end{align}
with slightly more stringent bounds when $U=0$ is assumed. Taking the central values as representative, one finds
\begin{align}
\delta\zeta_L\simeq 2\cdot 10^{-4};\quad \delta\zeta_R\simeq-1\cdot 10^{-4}
\end{align}
For the NNLO corrections, bounds are much weaker. At present, CDF~\cite{Aaltonen:2011zc}, CMS~\cite{Chatrchyan:2011rr} and D0~\cite{Abazov:2011qp} have reached the $10^{-2}$ precision on the nTGC, while {\mbox{ATLAS}}~\cite{Aad:2012awa,Aad:2013izg} provides the most precise determination to date, with bounds on all the parameters hovering around the lower $10^{-2}$ level. This is still far from the naive dimensional analysis estimate, which predicts effects at the $10^{-4}$ level. Translated into the EFT coefficients, naively one expects ${\tilde{c}}_{jW},{\tilde{c}}_{jB}\sim v^2/\Lambda^2\sim 7\cdot 10^{-3}$. Here I will use 
\begin{align}\label{cj}
\frac{{\tilde{c}}_{jW}}{\Lambda^2}\simeq 3\cdot 10^{-9}\,{\mathrm{GeV}}^{-2};\quad \frac{{\tilde{c}}_{jB}}{\Lambda^2}\simeq 2\cdot 10^{-9}\,{\mathrm{GeV}}^{-2}
\end{align}
which are on the conservative side. New physics scenarios with heavy fermions~\cite{Gounaris:2000tb,Dedes:2012me} are typical mechanisms to generate anomalous nTGV. Such contributions have to vanish at asymptotically large energies to comply with the triangle anomaly. However, if the heavy fermion thresholds sit not much above $3$ TeV, they can lead to sizeable enhancements on the NNLO coefficients. The plots of Fig.~\ref{fig:2} therefore show only a modest amount of new physics. If new physics appears at the TeV scale in the form of new fermion families, depending on their inter-family mass splittings, larger deviations can be easily generated.

The upper line in each panel of Fig.~\ref{fig:2} corresponds to the TT channel, which overwhelmingly dominates the cross section for both $\gamma Z$ and $ZZ$ production. The main source of new physics in the TT channel comes from oblique corrections, which correct the standard model vertices at the permille level. In contrast, new physics effects in the subdominant LT channel are parametrically enhanced, and even the very small corrections of Eq.~(\ref{cj}) become detectable at $\sqrt{s}\sim (0.6-1)$ TeV.   


\section{Comparison with $WW$ production}\label{sec:V}
It is instructive at this point to highlight the main differences between the EFT analysis presented here for neutral gauge boson pair production and that of Ref.~\cite{Buchalla:2013wpa} for the charged diboson production.
\begin{itemize}
\item In the charged case, the standard model contributions to triple gauge vertices appear first at tree level. In the neutral case, contributions arise at one-loop through (anomalous) fermionic triangles.
\item The number of independent triple gauge structures is substantially reduced in the neutral case and only comprises operators that either violate CP or P. Only the latter can interfere with the standard model contribution. This singles out 2 coefficients for $ZZ$ production and 4 for $\gamma Z$.  
\item In terms of effective operators, both processes are affected by the same set of NLO gauge-fermion and oblique corrections. However, triple gauge operators for the neutral case only appear at NNLO.  
\item Initial-state polarizations can be used in both cases to isolate different sets of EFT coefficients. However, in the charged case the $t$-channel is purely left-handed, while in the neutral case left and right-handed polarizations are diagrammatically equivalent.  
\item In the charged case, $s$-enhanced new physics contributions were generated by longitudinal (LL) final states. In the neutral case, LL final states are extremely suppressed and LT and TT final states are instead the dominant ones.  
\item Redundancies among operators can be used to relate NLO gauge-fermion, oblique and triple gauge operators. In $e^+e^-\to W^+W^-$, the $s$-enhanced triple gauge operators can be eliminated in favor of gauge-fermion ones. In the neutral case, the same relations can be used to eliminate the gauge-fermion operators in favor of the oblique S and T parameters.  
\item For charged diboson production, gauge-fermion operators parametrically decouple in $s$ from oblique and triple gauge operators. In the neutral case, this decoupling does not take place. Instead, oblique, gauge-fermion and triple gauge operators can be disentangled with final-state polarizations. 
\end{itemize}

\section{Conclusions}\label{sec:VI}

Given that so far there is no evidence of new physics below the TeV scale, adopting an effective field theory is the most general and efficient way to parametrize new physics effects at present-day hadron colliders and future linear colliders. In this paper I have presented a full-fledged EFT analysis of $ZZ$ and $\gamma Z$ production both for the linear and nonlinear realizations of electroweak symmetry breaking. The analysis includes all possible sources of new physics up to NNLO, which turn out to be parametrizable in terms of 6 parameters for $e^+e^-$ collisions and 8 for $pp$ collisions. For $e^+e^-$ collisions, all 6 parameters can be separately determined by exploiting the initial and final-state polarization structure of $ZZ$ and $\gamma Z$ production at typical projected linear collider energies, {\it{i.e.}}, $v^2\ll s\ll \Lambda^2$.  

A polarization analysis actually turns out to be extremely rewarding. While the TT channel is dominated by the standard model and almost saturates the cross section, the subdominant LT channel offers a clean window for new physics detection, where nTGV effects are dominant. In contrast, the LL channel turns out to be irrelevant. Notice that this phenomenological pattern is in sharp contrast with the one that emerges from the corresponding EFT analysis of $WW$ production~\cite{Buchalla:2013wpa}. 

The previous pattern shows that linear colliders are excellent laboratories to constrain nTGC. Using the magnifying power of linear collider energies in the LT channel, effects of NNLO coefficients can easily amount to $20\%$ corrections. Conversely, in the absence thereof, one can place very stringent bounds on the corresponding new physics operators.        


\section*{Acknowledgements}
I would like to thank Gerhard Buchalla for reading the draft and for the useful discussions that followed. This work was performed in the context of the ERC Advanced Grant project `FLAVOUR' (267104) and was supported in part by the 
DFG cluster of excellence `Origin and Structure of the Universe'.


\end{document}